\documentclass[journal,a4paper]{IEEEtran}
\usepackage[T1]{fontenc}
\usepackage{layouts}

\usepackage{cite}

\usepackage[pdftex,colorlinks=true,bookmarks=true,citecolor=blue,urlcolor=blue]{hyperref} 

\ifCLASSINFOpdf
  \usepackage[pdftex]{graphicx}
   \graphicspath{{./figs/}}
  \DeclareGraphicsExtensions{.pdf,.jpeg,.png}
\else
\fi
\usepackage{amsmath}
\usepackage{amsthm}
\usepackage[cmintegrals]{newtxmath}
\usepackage{bm}

\usepackage{algorithmic}

\usepackage{array}

\ifCLASSOPTIONcompsoc
  \usepackage[caption=false,font=normalsize,labelfont=sf,textfont=sf]{subfig}
\else
  \usepackage[caption=false,font=footnotesize]{subfig}
\fi
\usepackage{mathrsfs}  
\usepackage{amsbsy}

\newcommand{\field}[1]{\mathbb{#1}}
\newcommand{\fs}[1]{\mathsf{#1}}

\DeclareMathOperator{\Wrons}{\mathscr{W}}

\newcommand{\tp}{\intercal}

\newcommand{\ovl}[1]{\overline{#1}}

\newcommand{\bigO}[1]{\mathop{\mathcal{O}}\left(#1\right)}

\let\Re\relax
\DeclareMathOperator{\Re}{Re}
\let\Im\relax
\DeclareMathOperator{\Im}{Im}

\newcommand{\vs}[1]{\boldsymbol{#1}}

\newtheorem{rem}{Remark}[section]

\usepackage{enumitem}

\begin{document}
%
\title{Darboux Transformation: New Identities}
%
%
%


\author{Vishal Vaibhav
\thanks{Email:~\tt{vishal.vaibhav@gmail.com}}
}

\IEEEpubid{\begin{minipage}[t]{\textwidth}
\vskip2em
\centering
\copyright~2019 IEEE. Personal use of this material is permitted. Permission 
from IEEE must be obtained for all other uses, including reprinting/republishing this 
material for advertising or promotional purposes, collecting new collected works for 
resale or redistribution to servers or lists, or reuse of any copyrighted component 
of this work in other works.
\end{minipage}}


\maketitle

\begin{abstract}
This letter reports some new identities for multisoliton potentials that are
based on the explicit representation provided by the Darboux matrix. These
identities can be used to compute the complex gradient of the energy content of
the tail of the profile with respect to the discrete eigenvalues and the norming
constants. The associated derivatives are well defined in the framework of the so-called 
Wirtinger calculus which can aid a complex variable based optimization
procedure in order to generate multisolitonic signals with desired effective
temporal and spectral width.
\end{abstract}


%
\IEEEpeerreviewmaketitle

\section{Introduction}
This letter deals with the Darboux representation of multisoliton solutions of the 
nonlinear Schr\"odinger equation. As carriers of information, these multisolitonic 
signals offer a promising solution to the problem of nonlinear signal distortions 
in fiber optic channels~\cite{TPLWFK2017}. In any nonlinear Fourier transform (NFT) 
based transmission methodology seeking to modulate 
the discrete spectrum of the multisolitons, the unbounded support of such 
signals (as well as its Fourier spectrum) presents some challenges in achieving 
the best possible spectral efficiency~\cite{SABT2017,SABT2018ITG} which forms
part of the motivation for this work.

The Darboux transformation has proven to be an extremely powerful tool in
handling the discrete part of the nonlinear Fourier spectrum. The rational
structure of the associated Darboux matrix was recently exploited to obtain fast
inverse NFT algorithms in~\cite{V2017INFT1,V2018BL}. In the particular case of
$K$-soliton solutions, the rational structure of the Darboux matrix facilitates
the exact solution of the Zakharov-Shabat problem~\cite{AKNS1974} for (doubly-) truncated
version of the signal via the solution of an associated Riemann-Hilbert
problem~\cite{V2018CNSNS}. In~\cite{V2017INFT1,V2018CNSNS}, an exact method for
computing the energy content of the ``tails'' of $K$-soliton solutions was
reported which was again based on the Darboux transformation. This method was
further used in~\cite{V2018TL} to establish the sufficient conditions
for either one-sided or compact support of the signals resulting from
``addition'' of boundstates. Given that the complexity of 
computing the Darboux matrix coefficients is $\bigO{K^2}$, these methods turn
out to be extremely efficient eliminating any need for heuristic approaches based
on the asymptotic expansions (with respect to the windowing parameter, say, $\tau$). 
The present work, therefore, tries to further reinforce the idea that the
Darboux representation can potentially facilitate a number of design and signal 
processing aspects of $K$-soliton solutions. For the general case, when the
reflection coefficient is bandlimited, the work presented in~\cite{V2018BL1} may
allow us to compute the Jost solutions of the seed potential with extremely high
accuracy.

The present work is also motivated by the fact that the recent
attempts~\cite{SABT2017,SABT2018ITG,SABT2018}\footnote{The readers are warned
that the representation of the
norming constants used in these papers do not follow the standard convention
and are incorrect in many cases. The relationship $b_k=b(\zeta_k)$, where
$(\zeta_k,b_k)$ in the tuple comprising the discrete eigenvalue and the
corresponding norming constant, does not hold when $b$ does not have an analytic
continuation in $\field{C}_+$. In fact, from~\cite{Lin1990} it is known that
the $b$ remains invariant when boundstates are added to an arbitrary profile so
that $b_k\neq b(\zeta_k)$ even if $b(\zeta_k)$ exists.} towards optimizing 
the generated multisolitonic signals are either based on brute-force methods or 
asymptotic expansions with respect to the windowing parameter. These methods have serious 
drawbacks either because they do not scale well in complexity when the number of boundstates 
are only moderately high or because they are not reliable in the absence of a prior 
knowledge of the goodness of the approximations made. In this letter, we present
some new identities that can potentially make the optimization problem amenable
to some of the powerful optimization procedures available in the literature
(see~\cite{SBL2012} and the references therein) at
the same time completely circumventing the need for any heuristics. Given that
the independent variables (i.e. discrete eigenvalues and the norming constants) in 
the optimization procedure are complex in nature, the framework based on
Wirtinger calculus presented in~\cite{SBL2012} appears to be more convenient.

The main contributions of this work are presented in~Sec.~\ref{sec:TW} which
deals with the temporal width which is followed by a brief discussion of
estimation of spectral width in Sec.~\ref{sec:SW}. The letter concludes with some
examples in Sec.~\ref{sec:examples} where calculation can be carried out in a simple manner.

\section{Temporal Width}\label{sec:TW}
The temporal width a $K$-soliton solution can be defined via the
$\fs{L}^2$-norm of the profile which is also related to the energy of the pulse.
Let the energy content of the ``tails'' of the profile, denoted by
$\mathcal{E}^{(\pm)}(t)$, be defined by
\begin{equation}
\mathcal{E}^{(-)}(t)=\int^{t}_{-\infty}|q(s)|^2ds,\quad
\mathcal{E}^{(+)}(t)=\int_{t}^{\infty}|q(s)|^2ds,
\end{equation}
so that
\begin{equation}
\mathcal{E}(t)=\left[\mathcal{E}^{(-)}(-t)+\mathcal{E}^{(+)}(t)\right],
\end{equation}
characterizes the total energy in the tails which is a fraction of the total energy of a 
$K$-soliton solution. The total energy is given by $\|q\|_2^2=4\sum_{k=1}^{K}\Im{\zeta_k}$. Now, if the 
tolerance for the fraction of total energy in the tails is 
$\epsilon_{\text{tails}}$, then effective the
support, $[-\tau,\tau]$, of the profile must be chosen such that 
$\mathcal{E}(\tau)\leq\epsilon_{\text{tails}}\|q\|_2^2$.
Therefore, the effective temporal width of the $K$-soliton solution can then be defined as 
$2\tau$ which is a function of $\epsilon_{\text{tails}}$. 

For the specific case of $K$-soliton solutions, an exact recipe for computing 
$\mathcal{E}(\tau)$ was reported in~\cite{V2017INFT1,V2018CNSNS} which we
summarize briefly as follows: Let ${v}(t;\zeta) = (\vs{\phi},\vs{\psi})$ be 
the matrix form of the Jost solutions. Then, from the standard
theory of scattering transforms~\cite{AKNS1974}, it is known that, for
$\zeta\in\ovl{\field{C}}_+$,
\begin{equation}\label{eq:asymp-Jost}
{v}_Ke^{i\sigma_3\zeta t}=
\begin{pmatrix}
 1+\frac{1}{2i\zeta}\mathcal{E}^{(-)}& \frac{1}{2i\zeta}q(t)\\
-\frac{1}{2i\zeta}r(t)& 1+\frac{1}{2i\zeta}\mathcal{E}^{(+)}
\end{pmatrix}+\bigO{\frac{1}{\zeta^2}}.
\end{equation}
Now, the $K$-soliton potentials along with their Jost solutions can be computed quite 
easily using the \emph{Darboux transformation} (DT)~\cite{V2017INFT1,V2018CNSNS}. Let $\mathfrak{S}_K$ 
be the discrete spectrum of a $K$-soliton potential. The seed solution here 
corresponds to the \emph{null} potential; therefore, 
$v_0(t;\zeta)=e^{-i\sigma_3\zeta t}$.
The augmented matrix Jost solution ${v}_K(t;\zeta)$ can be obtained from the 
seed solution $v_0(t;\zeta)$ using the Darboux matrix as 
${v}_K(t;\zeta)=\mu_{K}(\zeta)D_{K}(t;\zeta,\mathfrak{S}_K)v_0(t;\zeta)$ for 
$\zeta\in\ovl{\field{C}}_+$. The Darboux transformation can be implemented 
as a recursive scheme. Let us define the successive discrete spectra
$\emptyset=\mathfrak{S}_0\subset\mathfrak{S}_1\subset\mathfrak{S}_2
\subset\ldots\subset\mathfrak{S}_K$ such that 
${\mathfrak{S}}_j=\{(\zeta_j,b_j)\}\cup{\mathfrak{S}}_{j-1}$ for
$j=1,2,\ldots,K$ where $(\zeta_j,b_j)$ are distinct elements of 
$\mathfrak{S}_K$. The Darboux matrix of degree $K>1$ can be factorized into
Darboux matrices of degree one as
\begin{multline}\label{eq:DT-mat}
D_K(t;\zeta,\mathfrak{S}_K|\mathfrak{S}_0)
=D_1(t;\zeta,\mathfrak{S}_{K}|\mathfrak{S}_{K-1})\\\times
D_1(t;\zeta,\mathfrak{S}_{K-1}|\mathfrak{S}_{K-2})\times
\ldots\times D_1(t;\zeta,\mathfrak{S}_1|\mathfrak{S}_0),
\end{multline}
where $D_1(t;\zeta,\mathfrak{S}_{j}|\mathfrak{S}_{j-1}),\,j=1,\ldots,K$ are 
the successive Darboux matrices of degree 
one with the convention that 
$(\zeta_{j},b_{j})=\mathfrak{S}_{j}\cap\mathfrak{S}_{j-1}$ is the bound
state being added to the seed potential whose discrete spectra is
$\mathfrak{S}_{j-1}$. Note that the Darboux matrices of degree one can be
stated as
\begin{equation}
D_1(t;\zeta,\mathfrak{S}_{j}|\mathfrak{S}_{j-1})= \zeta \sigma_0-
\begin{pmatrix}
\frac{|\beta_{j-1}|^2\zeta_j+\zeta_j^*}{1+|\beta_{j-1}|^2} 
&\frac{(\zeta_j-\zeta_j^*)\beta_{j-1}}{1+|\beta_{j-1}|^2}\\
\frac{(\zeta_j-\zeta_j^*)\beta^*_{j-1}}{1+|\beta_{j-1}|^2}
&\frac{\zeta_j+\zeta_j^*|\beta_{j-1}|^2}{1+|\beta_{j-1}|^2}
\end{pmatrix},
\end{equation}
and
\begin{equation}
\beta_{j-1}(t;\zeta_j, b_j) =
\frac{\phi_1^{(j-1)}(t;\zeta_j)-b_{j}\psi_1^{(j-1)}(t;\zeta_j)}
{\phi_2^{(j-1)}(t;\zeta_j) - b_{j}\psi_2^{(j-1)}(t;\zeta_j)},
\end{equation}
for $(\zeta_j,b_j)\in\mathfrak{S}_K$ and the successive Jost solutions, 
${v}_{j} = (\vs{\phi}_{j},\vs{\psi}_{j})$, needed in this ratio are computed as
\begin{equation}
{v}_j(t;\zeta)=(\zeta-\zeta^*_j)^{-1}D_{1}(t;\zeta,\mathfrak{S}_j|\mathfrak{S}_{j-1})
v_{j-1}(t;\zeta).
\end{equation}
The potential is given by 
\begin{equation}\label{DT-iter-pot}
q_j = q_{j-1} -
2i\frac{(\zeta_j-\zeta_j^*)\beta_{j-1}}{1+|\beta_{j-1}|^2}.
\end{equation}
and
\begin{equation}\label{eq:energy-iter}
\mathcal{E}^{(-)}_j =\mathcal{E}^{(-)}_{j-1}
+\frac{4\Im(\zeta_j)}{1+|\beta_{j-1}|^{-2}},\quad
\mathcal{E}^{(+)}_j =\mathcal{E}^{(+)}_{j-1}
+\frac{4\Im(\zeta_j)}{1+|\beta_{j-1}|^{2}},
\end{equation}
so that $\mathcal{E}^{(\mp)}(t)=\mathcal{E}_K^{(\mp)}(t)$ with
$\mathcal{E}_0^{(\mp)}(t)\equiv0$. Next, our objective is to 
compute the derivatives of $\mathcal{E}(\tau)$ with
respect to the discrete spectra of $K$-soliton in order to facilitate
optimization procedures that are based on gradients.
In the following, we use the notation
\begin{equation}
\partial_{Z_j}       = \frac{\partial}{\partial Z_j},\quad
\ovl{\partial}_{Z_j} = \frac{\partial}{\partial ( Z^*_j)}.
\end{equation}
For real-valued function $f$, it is known that 
$\left(\partial_{Z_j} f\right)^*= \ovl{\partial}_{Z_j}f$. In Wirtinger calculus,
the complex gradient is defined as 
$(\partial_{Z_1},\ldots\partial_{Z_K},\ovl{\partial}_{Z_1},\ldots,\ovl{\partial}_{Z_K})^{\tp}$
\begin{rem}
Here, we have described the Darboux transformation only for the case of
$K$-soliton potentials; however, the recipe can be easily adapted to the case of
arbitrary seed potentials. Note that this would require explicit knowledge of
$v_0(t;\zeta)$ and $\mathcal{E}_0^{(\mp)}(t)$.
\end{rem}
\subsection{Derivatives with respect to norming constants}
Note that $\mathcal{E}^{(-)}_{K-1}(-\tau)$ and $\mathcal{E}^{(+)}_{K-1}(\tau)$ are
independent of $b_K$; therefore,
\begin{equation}
\partial_{b_K}\mathcal{E}^{(\mp)}(\mp\tau)=
\mp4\Im(\zeta_K)\partial_{b_K}\left[1+|\beta_{K-1}(\mp\tau)|^{2}\right]^{-1}.
\end{equation}
By direct calculation, we have 
\begin{equation}\label{eq:del-bK}
\partial_{b_K}\beta_{K-1}(t;\zeta_K, b_K)
=\frac{a_{K-1}(\zeta_{K})}
{\left[\phi_2^{(K-1)} - b_{K}\psi_2^{(K-1)}\right]^2(t;\zeta_K)},
\end{equation}
where we have used to the Wronskian relation
\begin{equation}
\begin{split}
a_{K-1}(\zeta)&=\Wrons\left(\vs{\phi}_{K-1},\vs{\psi}_{K-1}\right)
=\prod_{k=1}^{K-1}\left(\frac{\zeta-\zeta_k}{\zeta-\zeta^*_k}\right).
\end{split}
\end{equation}
Using the identity~\eqref{eq:del-bK}, it is straightforward to work out:
\begin{multline}
\partial_{b_K}\mathcal{E}^{(\mp)}(\mp\tau)
=\pm4\Im(\zeta_K)\frac{|\beta_{K-1}|^{2}}{\left[1+|\beta_{K-1}|^{2}\right]^2}\\
\times\left.\frac{a_{K-1}(\zeta_{K})}
{\left[\phi_1^{(K-1)} - b_{K}\psi_1^{(K-1)}\right]
\left[\phi_2^{(K-1)} - b_{K}\psi_2^{(K-1)}\right]}\right|_{t=\mp\tau}.
\end{multline}
Note that $\zeta_K$ is the last eigenvalue to be added using the DT iterations.
Given that there is no restriction on the order in which the eigenvalues can 
be added, we can always choose $\zeta_k$ to be added last. This would determine 
$\partial_{b_k}\mathcal{E}^{(\pm)}$ using DT iterations for arbitrary $k$. Thus,
the complexity of computing $K$ derivatives works out to be $\bigO{K^3}$.

Before we conclude this discussion, let us examine the case of multisoliton
solutions when $\tau$ is large. In this limit, we have 
\begin{equation}
\beta_{K-1}(-\tau)\sim\frac{b^{-1}_Ke^{2i\zeta_K \tau}}{a_{K-1}(\zeta_K)},\quad
\beta^{-1}_{K-1}(+\tau)\sim\frac{b_Ke^{2i\zeta_K \tau}}{a_{K-1}(\zeta_K)},
\end{equation}
so that
\begin{equation}
\partial_{b_K}\mathcal{E}^{(\mp)}(\mp\tau)\sim\mp\frac{4\Im(\zeta_K)}{b_{K}}|\beta_{K-1}(\mp\tau)|^{\pm2}.
\end{equation}
Thus, the stationary condition $\partial_{b_K}\mathcal{E}(\tau)=0$ translates into
$|b_K|=1$. Therefore, asymptotically, $|b_j|=1$ minimizes the energy in the
tails. This result can be easily verified from~\eqref{eq:energy-iter} which in
the limit of large $\tau$ gives 
\begin{multline}
\mathcal{E}(\tau)
\sim\sum_{j=1}^K\frac{4\Im(\zeta_j)}{|a_{j-1}(\zeta_j)|^2}
\left(|b_j|^2+\frac{1}{|b_j|^2}\right)e^{-4\eta_j \tau}\\
\leq\sum_{j=1}^K\frac{8\Im(\zeta_j)}{|a_{j-1}(\zeta_j)|^2}e^{-4\eta_j \tau}.
\end{multline}

\subsection{Derivatives with respect to discrete eigenvalues}
Let ${V}_j(t;\zeta)=(\zeta-\zeta^*_j){v}_j(t;\zeta)$ and define ${V}_{j} =
(\vs{\Phi}_{j},\vs{\Psi}_{j})$ so that
\begin{equation}
{V}_j(t;\zeta)=D_{1}(t;\zeta,\mathfrak{S}_j|\mathfrak{S}_{j-1})
V_{j-1}(t;\zeta).
\end{equation}
The ratio $\beta_j$ can also be computed in
terms of the modified Jost solutions on account of the fact that
$(\zeta-\zeta_j)$ falls out of the equation while taking the ratio:
\begin{equation}
\beta_{j-1}(t;\zeta_j, b_j) =
\frac{\Phi_1^{(j-1)}(t;\zeta_j)-b_{j}\Psi_1^{(j-1)}(t;\zeta_j)}
{\Phi_2^{(j-1)}(t;\zeta_j) - b_{j}\Psi_2^{(j-1)}(t;\zeta_j)},
\end{equation}
This gives us the opportunity to compute the derivatives with respect to
$\zeta$ recursively:
\begin{equation}
\partial_{\zeta}{V}_j(t;\zeta)=V_{j-1}(t;\zeta)+D_{1}(t;\zeta,\mathfrak{S}_j|\mathfrak{S}_{j-1})
\partial_{\zeta}V_{j-1}(t;\zeta).
\end{equation}
Using the notation
$\Wrons_{\zeta}\left(u,v\right)=\left(u\partial_{\zeta}v-v\partial_{\zeta}u\right)$ for the
Wronskian of scalar functions, let us introduce
\begin{equation}
\begin{split}
&W^{(K-1)}_1(t;\zeta_K)=\Wrons_{\zeta}\left(\Phi_2^{(K-1)},\Phi_1^{(K-1)}\right)(t;\zeta_K),\\
&W^{(K-1)}_2(t;\zeta_K)=\Wrons_{\zeta}\left(\Psi_2^{(K-1)},\Psi_1^{(K-1)}\right)(t;\zeta_K).
\end{split}
\end{equation}
By direct calculation, we have
\begin{multline}
\partial_{\zeta_K}\beta_{K-1}(t;\zeta_K, b_K)
=\frac{\left[W^{(K-1)}_1+b^2_KW^{(K-1)}_2\right]}
{\left[\Phi_2^{(K-1)}-b_{K}\Psi_2^{(K-1)}\right]^2}(t;\zeta_K)\\
-\frac{b_K\partial_{\zeta}a_{K-1}(\zeta_K)}
{\left[\Phi_2^{(K-1)}-b_{K}\Psi_2^{(K-1)}\right]^2(t;\zeta_K)}.
\end{multline}
Note that $\mathcal{E}^{(-)}_{K-1}(-\tau)$ and $\mathcal{E}^{(+)}_{K-1}(\tau)$ are
independent of $\zeta_K$; therefore, using the above identity, it is straightforward to obtain 
\begin{multline}
\partial_{\zeta_K}\mathcal{E}^{(\mp)}(\mp\tau)
=\frac{2}{1+|\beta_{K-1}|^{\mp2}}
+\frac{4\Im(\zeta_K)|\beta_{K-1}|^{2}}{\left[1+|\beta_{K-1}|^{2}\right]^2}\times\\
\left[\frac{W^{(K-1)}_1
-b_K\partial_{\zeta}a_{K-1}(\zeta_K)
+b^2_KW^{(K-1)}_2}
{\left(\Phi_1^{(K-1)} - b_{K}\Psi_1^{(K-1)}\right)
\left(\Phi_2^{(K-1)} - b_{K}\Psi_2^{(K-1)}\right)}\right]_{t=\mp\tau}.
\end{multline}
Following as in the case of norming constants, we can always choose 
$\zeta_k$ to be added last so that $\partial_{\zeta_k}\mathcal{E}^{(\pm)}$ can
be determined using DT iterations for arbitrary $k$. Thus, the complexity of 
computing $K$ derivatives again works out to be $\bigO{K^3}$.

\section{Spectral Width}\label{sec:SW}
Consider the Fourier spectrum of the multisoliton potential denoted by
$Q(\xi)=\int q(t)e^{-i\xi t}dt,\,\xi\in\field{R}$. Let us observe that 
the following quantities can be expressed entirely in terms 
of the discrete eigenvalues: 
\begin{equation}
\left\{\begin{aligned}
C_1&=-\int q^*(\partial_tq)dt=4i\sum_{k}\Im{\zeta^2_k},\\
C_2&=\int\left(|q|^4-|\partial_tq|^2\right)dt
=-\frac{16}{3}\sum_{k}\Im{\zeta^3_k}.
\end{aligned}\right.
\end{equation}
with $C_0=\|q\|_2^2$. These quantities do not evolve as the pulse
propagates along the fiber. From~\cite{V2018CNSNS}, the variance 
$\langle\Delta\xi^2\rangle$ is given by
\begin{equation}
\langle\Delta\xi^2\rangle
=\frac{\int |q|^4dt}{C_0}+\frac{C_1^2}{C_0^2}-\frac{C_2}{C_0}
\leq \|q\|^2_{\infty}+\frac{C_1^2}{C_0^2}-\frac{C_2}{C_0}.
\end{equation}
This quantity characterizes the width of the Fourier spectrum. The biquadratic 
integral above cannot be computed exactly in general, however, $\|q\|_{\infty}$ can be computed
in a straightforward manner: From~\eqref{DT-iter-pot}, we have 
$\|q_j\|_{\infty} \leq \|q_{j-1}\|_{\infty}+2\Im(\zeta_j)$, we have 
$\|q_K\|_{\infty} \leq 2\sum_{k=1}^K\Im(\zeta_j)$ which 
yields $\langle\Delta\xi^2\rangle\leq S$ where (correcting a typographical
error in~\cite{V2018CNSNS})
\begin{equation}
S=\left(\frac{C^2_0}{4}+\frac{C_1^2}{C_0^2}-\frac{C_2}{C_0}\right).
\end{equation}
Note that this inequality holds irrespective of how the pulse evolves as it
propagates along the fiber. 
\section{Examples}\label{sec:examples}
\subsection{One-sided effective support}
Let us consider the case where we want to introduce a boundstate with eigenvalue $\zeta_1$ to any
arbitrary profile such that the energy content of the tail
$[\tau,\infty)\,(\tau>0)$ is 
$\mathcal{E}_0^{(+)}(\tau)$. The problem is to determine the norming constant $b_1$ which
minimizes $\mathcal{E}^{(+)}=\mathcal{E}^{(+)}_1(\tau)$. To this end, setting
$\partial_1\mathcal{E}^{(+)}(\tau)=0$, we have
\[
\left.{\left[\phi_1^{(0)} - b_{1}\psi_1^{(0)}\right]
\left[\phi_2^{(0)} - b_{1}\psi_2^{(0)}\right]}\right|_{t=\tau,\,\zeta=\zeta_1}=0,
\]
which yields
\[
b_{1}\in\left\{\frac{\phi_1^{(0)}(\tau;\zeta_1)}{\psi_1^{(0)}(\tau;\zeta_1)},
\frac{\phi_2^{(0)}(\tau;\zeta_1)}{\psi_2^{(0)}(\tau;\zeta_1)}\right\}.
\]
It is easy to verify that the first choice corresponds to maximum
$\mathcal{E}^{(+)}(\tau)$ which leaves us with $b_{1}={\phi_2^{(0)}(\tau;\zeta_1)}/{\psi_2^{(0)}(\tau;\zeta_1)}$
so that $\mathcal{E}^{(+)}(\tau)=\mathcal{E}_0^{(+)}(\tau)$, i.e., no part of the soliton's
energy goes into the tail $[\tau,\infty)$. By a recursive argument, the conclusion
holds for any number of boundstates provided 
$b_{j}={\phi_2^{(0)}(\tau;\zeta_j)}/{\psi_2^{(0)}(\tau;\zeta_j)}$.

\subsection{Adding a boundstate to a symmetric profile}
Let us consider the case where we want to introduce a boundstate with eigenvalue $\zeta_1$ to any
arbitrary seed profile. The energy content of the tail
$\field{R}\setminus(-\tau,\tau)$ of the seed profile is 
$\mathcal{E}_0(\tau)$. The problem is to determine the norming constant $b_1$ which
minimizes $\mathcal{E}(\tau)$. To this end, setting
$\partial_1\mathcal{E}(\tau)=0$, we have
\begin{multline}
\left.\frac{\left[\phi_1^{(0)} - b_{1}\psi_1^{(0)}\right]
\left[\phi_2^{(0)} - b_{1}\psi_2^{(0)}\right]}
{\left[|\phi_1^{(0)} - b_{1}\psi_1^{(0)}|^2+|\phi_2^{(0)} - b_{1}\psi_2^{(0)}|^{2}\right]^2}
\right|_{t=-\tau}\\
=\left.\frac{\left[\phi_1^{(0)} - b_{1}\psi_1^{(0)}\right]
\left[\phi_2^{(0)} - b_{1}\psi_2^{(0)}\right]}
{\left[|\phi_1^{(0)} - b_{1}\psi_1^{(0)}|^2+|\phi_2^{(0)} - b_{1}\psi_2^{(0)}|^{2}\right]^2}
\right|_{t=\tau}.
\end{multline}
For the sake of simplicity, we assume that the seed profile is symmetric.
Further, we also assume that $\zeta_1=i\eta_1$ so that
\begin{equation}
\left\{
\begin{aligned}
&\phi^{(0)}_1(-t;i\eta_1) = \psi^{(0)*}_2(t;i\eta_1),\\
&\phi^{(0)}_2(-t;i\eta_1) = \psi^{(0)*}_1(t;i\eta_1),
\end{aligned}\right.
\end{equation}
with $a(i\eta)=a^*(i\eta)$. In the following, we set $t=\tau$. Then, 
using the symmetry relations, we obtain  
\begin{multline}\label{eq:b1-symm}
\frac{\left[\phi_1^{(0)*} - (1/b_{1})\psi_1^{(0)*}\right]
\left[\phi_2^{(0)*} - (1/b_{1})\psi_2^{(0)*}\right]}
{\left[\phi_1^{(0)} - b_{1}\psi_1^{(0)}\right]
\left[\phi_2^{(0)} - b_{1}\psi_2^{(0)}\right]}\frac{b_1^2}{|b_1|^4}\\
=\frac{\left[|\phi_1^{(0)} - (1/b^*_{1})\psi_1^{(0)}|^2+|\phi_2^{(0)} -
(1/b^*_{1})\psi_2^{(0)}|^{2}\right]^2}
{\left[|\phi_1^{(0)} - b_{1}\psi_1^{(0)}|^2+|\phi_2^{(0)} -
b_{1}\psi_2^{(0)}|^{2}\right]^2}.
\end{multline}
Physically, $\log|b_1|$ is related to the translation of the profile; therefore, it
is easy to conclude, for a symmetrical profile, that the extrema is obtained for 
$|b_1|=1$. Putting 
\begin{equation}
\begin{split}
A &= i\left(\psi_1^{(0)}\psi_2^{(0)}-\phi_1^{(0)*}\phi_2^{(0)*}\right),\\
B &= i\left(\psi_1^{(0)}\phi_2^{(0)}-\psi_1^{(0)*}\phi_2^{(0)*}\right)\\
&= i\left(\phi_1^{(0)}\psi_2^{(0)}-\phi_1^{(0)*}\psi_2^{(0)*}\right),
\end{split}
\end{equation}
in~\eqref{eq:b1-symm} and using $|b_1|=1$, we have $Ab_1^2-2Bb_1+A^*=0$. The solution of 
this equation works out to be
\begin{equation}
b_1 
=\frac{B\pm i\sqrt{|A|^2-B^2}}{A}
=\frac{B\pm i\sqrt{\Delta}}{A}.
\end{equation}
From the relations
\begin{equation*}
\begin{split}
B^2&=2|\psi_1^{(0)}|^2|\phi_2^{(0)}|^2-2\Re[(\psi_1^{(0)}\phi_2^{(0)})^2]\\
&=2|\phi_1^{(0)}|^2|\psi_2^{(0)}|^2-2\Re[(\phi_1^{(0)}\psi_2^{(0)})^2],\\
|A|^2&=|\psi_1^{(0)}|^2|\psi_2^{(0)}|^2+|\phi_1^{(0)}|^2|\phi_2^{(0)}|^2
-2\Re[\psi_1^{(0)}\psi_2^{(0)}\phi_1^{(0)}\phi_2^{(0)}],
\end{split}
\end{equation*}
we have
\begin{multline}
\Delta=[a(i\eta_1)]^2+|\psi_1^{(0)}|^2|\psi_2^{(0)}|^2+|\phi_1^{(0)}|^2|\phi_2^{(0)}|^2\\
-|\phi_1^{(0)}|^2|\psi_2^{(0)}|^2-|\psi_1^{(0)}|^2|\phi_2^{(0)}|^2.
\end{multline}
In order to show that $\Delta\geq0$, consider
\begin{equation*}
\begin{split}
\Delta&=(|\psi_1^{(0)}\psi_2^{(0)}|-|\phi_1^{(0)}\phi_2^{(0)}|)^2\\
&\qquad+\left|\phi_1^{(0)}\psi_2^{(0)}-\psi_1^{(0)}\phi_2^{(0)}\right|^2
-\left||\phi_1^{(0)}\psi_2^{(0)}|-|\psi_1^{(0)}\phi_2^{(0)}|\right|^2\\
&=(|\psi_1^{(0)}\psi_2^{(0)}|-|\phi_1^{(0)}\phi_2^{(0)}|)^2\\
&\qquad+\left(\left|\phi_1^{(0)}\psi_2^{(0)}-\psi_1^{(0)}\phi_2^{(0)}\right|
-\left||\phi_1^{(0)}\psi_2^{(0)}|-|\psi_1^{(0)}\phi_2^{(0)}|\right|\right)\\
&\qquad\times\left(\left|\phi_1^{(0)}\psi_2^{(0)}-\psi_1^{(0)}\phi_2^{(0)}\right|
+\left||\phi_1^{(0)}\psi_2^{(0)}|-|\psi_1^{(0)}\phi_2^{(0)}|\right|\right),
\end{split}
\end{equation*}
which shows that $\Delta\geq0$. Therefore, the extremal points for $b_1$ are
given by
\begin{equation}
\arg b_1 
=\pm\arg\left[\frac{B+i\sqrt{\Delta}}{|A|}\right]-\arg A.
\end{equation}

\subsubsection{Symmetric $2$-soliton}
The general result derived above can be applied to a symmetric $2$-soliton potential. Let 
us assume that the seed potential is a symmetric $1$-soliton potential with the discrete spectrum given
by $\{(i\eta_0, e^{i\theta_0})\}$. The boundstate being introduced is
characterized by $(i\eta_1, e^{i\theta_1})$. 
Expression for the Jost solutions can be obtained from~\eqref{eq:DT-mat} which
leads to $B=0$ %
so that $b_1=\pm e^{i\theta_0}$. It can be directly verified that
$b_1=e^{i\theta_0}$ corresponds to the minima of $\mathcal{E}_1(\tau)$ for all
$\tau>0$ as follows: Given the symmetric nature of the profile, it suffices to 
find the minima of $\mathcal{E}_1^{(+)}(\tau)$ which reads as
\begin{equation}
\mathcal{E}_1^{(+)}(\tau)=\mathcal{E}_0^{(+)}(\tau)
+\frac{4\eta_1Y^{-1}}{X^{-1}+Y^{-1}}\left(\frac{H}{G}\right)
\left(\frac{1-G\cos\theta}{1-H\cos\theta}\right),
\end{equation}
where $\theta=\theta_1-\theta_0$ and
\begin{equation*}
\left\{\begin{aligned}
X^{-1}&=e^{-2\eta_0\tau}+a_0(i\eta_1)e^{2\eta_0\tau}\\
&=\frac{2\eta_1\cosh(2\eta_0\tau)}{\eta_0+\eta_1}\left[1-\frac{\eta_0}{\eta_1}\tanh(2\eta_0\tau)\right],\\
Y^{-1}&=a_0(i\eta_1)e^{-2\eta_0\tau}+e^{2\eta_0\tau}\\
&=\frac{2\eta_1\cosh(2\eta_0\tau)}{\eta_0+\eta_1}\left[1+\frac{\eta_0}{\eta_1}\tanh(2\eta_0\tau)\right],\\
G&=\frac{2}{Ye^{2\eta_1\tau}+Y^{-1}e^{-2\eta_1\tau}},\\
H&=\frac{2(X^{-1}+Y^{-1})}{\left[(1+X^{-2})e^{2\eta_1\tau}+(1+Y^{-2})e^{-2\eta_1 \tau}\right]},
\end{aligned}\right.
\end{equation*}
It is straightforward to show that $0<G,H\leq1$, $Y^{-1}>0$ and $X^{-1}+Y^{-1}>0$. Now, from
\[
\frac{1-G\cos\theta}{1-H\cos\theta}
=\frac{1-G/H}{1-H\cos\theta}+\frac{G}{H},
\]
and
\begin{align*}
1-\frac{G}{H}
&=\frac{-\frac{(\eta_1-\eta_0)^2}{2\eta_1(\eta_1+\eta_0)}}{[Y^{-2}e^{-2\eta_1\tau}+e^{2\eta_1\tau}]}
\frac{\left[1+2\cosh(4\eta_1\tau)\right]}{\cosh(2\eta_1\tau)}\times\\
&\qquad{\left[\frac{\sinh[2(\eta_0+\eta_1)\tau]}{\eta_1+\eta_0}
+\frac{\sinh[2(\eta_1-\eta_0)\tau]}{\eta_1-\eta_0}\right]},
\end{align*}
it follows that $(1-G/H)\leq0$; therefore, the minima of
$\mathcal{E}_1^{(+)}(\tau)$ occurs at $\theta=2n\pi$ or $b_1=e^{i\theta_0}$.
\providecommand{\noopsort}[1]{}\providecommand{\singleletter}[1]{#1}%


\begin{thebibliography}{10}
\providecommand{\url}[1]{#1}
\csname url@samestyle\endcsname
\providecommand{\newblock}{\relax}
\providecommand{\bibinfo}[2]{#2}
\providecommand{\BIBentrySTDinterwordspacing}{\spaceskip=0pt\relax}
\providecommand{\BIBentryALTinterwordstretchfactor}{4}
\providecommand{\BIBentryALTinterwordspacing}{\spaceskip=\fontdimen2\font plus
\BIBentryALTinterwordstretchfactor\fontdimen3\font minus
  \fontdimen4\font\relax}
\providecommand{\BIBforeignlanguage}[2]{{%
\expandafter\ifx\csname l@#1\endcsname\relax
\typeout{** WARNING: IEEEtran.bst: No hyphenation pattern has been}%
\typeout{** loaded for the language `#1'. Using the pattern for}%
\typeout{** the default language instead.}%
\else
\language=\csname l@#1\endcsname
\fi
#2}}
\providecommand{\BIBdecl}{\relax}
\BIBdecl

\bibitem{TPLWFK2017}
S.~K. Turitsyn, J.~E. Prilepsky, S.~T. Le, S.~Wahls, L.~L. Frumin, M.~Kamalian,
  and S.~A. Derevyanko, ``Nonlinear {F}ourier transform for optical data
  processing and transmission: advances and perspectives,'' \emph{Optica},
  vol.~4, no.~3, pp. 307--322, Mar 2017.

\bibitem{SABT2017}
A.~Span, V.~Aref, H.~B{\"u}low, and S.~Ten~Brink, ``On time-bandwidth product
  of multi-soliton pulses,'' in \emph{Information Theory (ISIT), 2017 IEEE
  International Symposium on}.\hskip 1em plus 0.5em minus 0.4em\relax IEEE,
  2017, pp. 61--65.

\bibitem{SABT2018ITG}
------, ``Optimization of multi-soliton joint phase modulation for reducing the
  time-bandwidth product,'' in \emph{Photonic Networks; 19th
  ITG-Symposium}.\hskip 1em plus 0.5em minus 0.4em\relax VDE, 2018, pp. 1--8.

\bibitem{V2017INFT1}
V.~Vaibhav, ``Fast inverse nonlinear {F}ourier transformation using exponential
  one-step methods: {D}arboux transformation,'' \emph{Phys. Rev. E}, vol.~96,
  p. 063302, 2017.

\bibitem{V2018BL}
------, ``Fast inverse nonlinear {F}ourier transform,'' \emph{Phys. Rev. E},
  vol.~98, p. 013304, 2018.

\bibitem{AKNS1974}
M.~J. Ablowitz, D.~J. Kaup, A.~C. Newell, and H.~Segur, ``The inverse
  scattering transform - {F}ourier analysis for nonlinear problems,''
  \emph{Stud. Appl. Math.}, vol.~53, no.~4, pp. 249--315, 1974.

\bibitem{V2018CNSNS}
V.~Vaibhav, ``Exact solution of the {Z}akharov--{S}habat scattering problem for
  doubly-truncated multisoliton potentials,'' \emph{Commun. Nonlinear Sci.
  Numer. Simul.}, vol.~61, pp. 22--36, 2018.

\bibitem{V2018TL}
------, ``Nonlinear {F}ourier transform of time-limited and one-sided
  signals,'' \emph{J. Phys. A: Math. Theor.}, vol.~51, no.~42, p. 425201, 2018.

\bibitem{V2018BL1}
\BIBentryALTinterwordspacing
------. (2018) Nonlinearly bandlimited signals. [Online]. Available:
  \url{http://arxiv.org/abs/1811.06338}
\BIBentrySTDinterwordspacing

\bibitem{SABT2018}
\BIBentryALTinterwordspacing
A.~Span, V.~Aref, H.~B{\"u}low, and S.~Ten~Brink. (2018) Time-bandwidth product
  perspective for multi-soliton phase modulation. [Online]. Available:
  \url{http://arxiv.org/abs/1812.04443}
\BIBentrySTDinterwordspacing

\bibitem{Lin1990}
J.~Lin, ``Evolution of the scattering data under the classical {D}arboux
  transform for {SU}(2) soliton systems,'' \emph{Acta Mathematicae Applicatae
  Sinica}, vol.~6, no.~4, pp. 308--316, 1990.

\bibitem{SBL2012}
L.~Sorber, M.~van Barel, and L.~De~Lathauwer, ``Unconstrained optimization of
  real functions in complex variables,'' \emph{SIAM J. Optim.}, vol.~22, no.~3,
  pp. 879--898, 2012.

\end{thebibliography}
\end{document}